\documentclass[10pt]{article}

\usepackage{subfig}
\usepackage{graphicx}
\usepackage[colorlinks=true]{hyperref}
\newcommand\authormark[1]{\textsuperscript{#1}}
\newcommand\address[1]{%
	{\raggedright\small\itshape #1\par}
}
\begin{document}
\renewcommand{\arraystretch}{2}
\title{Simulating optical coherence tomography for observing nerve activity: a finite difference time domain bi-dimensional model.}

\author{F. Troiani,\authormark{1,*} K. Nikolic,\authormark{1} and T. G. Constandinou\authormark{1}}

\date{}
\maketitle
\address{\authormark{1}Centre for Bio-Inspired Technology, Imperial College London, Exhibition Road SW7 2AZ London, United Kingdom}

\textcolor{blue}{*\raggedright\footnotesize\itshape{f.troiani14@imperial.ac.uk}} 

\begin{abstract}
We present a finite difference time domain (FDTD) model for computation of A line scans in time domain optical coherence tomography (OCT). By simulating only the end of the two arms of the interferometer and computing the interference signal in post processing, it is possible to reduce the computational time required by the simulations and, thus, to simulate much bigger environments. Moreover, it is possible to simulate successive A lines and thus obtaining a cross section of the sample considered. In this paper we present the model applied to two different samples: a glass rod filled with water-sucrose solution at different concentrations and a peripheral nerve. This work demonstrates the feasibility of using OCT for non-invasive, direct optical monitoring of peripheral nerve activity, which is a long-sought goal of neuroscience.
\end{abstract}

\bibliographystyle{plain}
\bibliography{References}

\begin{thebibliography}{10}

\bibitem{McGill}
The mcgill physiology virtual lab - cap.
\newblock
  \url{http://www.medicine.mcgill.ca/physio/vlab/other_exps/CAP/nerve_anat.htm}.

\bibitem{SugarBrix}
United states department of agriculture sucrose conversion table, 1981.

\bibitem{Boppart2003}
Stephen~A. Boppart.
\newblock Optical coherence tomography: Technology and applications for
  neuroimaging.
\newblock {\em Psychophysiology}, 40(4):529--541, 2003.

\bibitem{Boppart1996}
Stephen~A. Boppart, Brett~E. Bouma, Mark~E. Brezinski, Guillermo~J. Tearney,
  and James~G. Fujimoto.
\newblock Imaging developing neural morphology using optical coherence
  tomography.
\newblock {\em Journal of Neuroscience Methods}, 70(1):65 -- 72, 1996.

\bibitem{Cohen1968}
Laurence~B. Cohen, Rrichard~D. Keynes, and Bertil Hille.
\newblock Light scattering and birefringence changes during nerve activity.
\newblock {\em Nature}, 218:438--441, 1968.

\bibitem{Hale1973}
George~M. Hale and Marvin~R. Querry.
\newblock Optical constants of water in the 200-nm to 200-$\mu$m wavelength
  region.
\newblock {\em Appl. Opt.}, 12(3):555--563, Mar 1973.

\bibitem{Huang1991}
David Huang, Eric~A. Swanson, Charles~P. Lin, Joel~S. Schuman, William~G.
  Stinson, Warren Chang, Michael~R. Hee, Thomas Flotte, Kenton Gregory,
  Carmen~A. Puliafito, and James~G. Fujimoto.
\newblock Optical coherence tomography.
\newblock {\em Science}, 254(5035):1178--1181, 1991.

\bibitem{Hubing2008}
T.~Hubing, C.~Su, H.~Zeng, and H.~Ke.
\newblock Survey of current computational electromagnetics techniques and
  software.
\newblock Technical report, Clemson University, 2008.

\bibitem{Kirillin2010}
Mikhail Kirillin, Igor Meglinski, Vladimir Kuzmin, Ekaterina Sergeeva, and
  Risto Myllyl\"{a}.
\newblock Simulation of optical coherence tomography images by monte carlo
  modeling based on polarization vector approach.
\newblock {\em Opt. Express}, 18(21):21714--21724, Oct 2010.

\bibitem{Magnain2015}
Caroline Magnain, Jean~C. Augustinack, Ender Konukoglu, Matthew~P. Frosch, Sava
  Sakad\^{z}i\'{c}, Ani Varjabedian, Nathalie Garcia, Van~J. Wedeen, David~A.
  Boas, and Bruce Fischl.
\newblock Optical coherence tomography visualizes neurons in human entorhinal
  cortex.
\newblock {\em Neurophotonics}, 2(1):015004, 2015.

\bibitem{Munro2016}
Peter~R.T. Munro.
\newblock Three-dimensional full wave model of image formation in optical
  coherence tomography.
\newblock {\em Opt. Express}, 24(23):27016--27031, Nov 2016.

\bibitem{Munro2015}
Peter~R.T. Munro, Andrea Curatolo, and David~D. Sampson.
\newblock Full wave model of image formation in optical coherence tomography
  applicable to general samples.
\newblock {\em Opt. Express}, 23(3):2541--2556, Feb 2015.

\bibitem{Periyasamy2016}
Vijitha Periyasamy and Manojit Pramanik.
\newblock Importance sampling-based monte carlo simulation of time-domain
  optical coherence tomography with embedded objects.
\newblock {\em Appl. Opt.}, 55(11):2921--2929, Apr 2016.

\bibitem{Schneider2010}
John~B. Schneider.
\newblock Understanding the finite-difference time-domain method, 2010.

\bibitem{Taflove2005}
Allen Taflove and Susan~C. Hagness.
\newblock {\em Computational Electrodynamics: The Finite-Difference Time-Domain
  Method, Third Edition}.
\newblock Artech House, 2005.

\bibitem{Troiani2016}
Francesca Troiani, Konstantin Nikolic, and Timothy~G Constandinou.
\newblock Optical coherence tomography for detection of compound action
  potential in xenopus laevis sciatic nerve, 2016.

\bibitem{Wilson1990}
B.C. Wilson and Steven~L. Jacques.
\newblock Optical reflectance and transmittance of tissues: principles and
  applications.
\newblock {\em IEEE Journal of Quantum Electronics}, 26(12):2186--2199, Dec
  1990.

\bibitem{Yee1966}
Kane~S. Yee.
\newblock Numerical solution of initial boundary value problems involving
  maxwell's equations in isotropic media.
\newblock {\em IEEE Trans. Antennas and Propagation}, pages 302--307, 1966.

\end{thebibliography}

\section{Introduction}
Optical coherence tomography is a low coherence interferometric technique that has been first used in 1991 to examine the peripapillary region of the retina\cite{Huang1991} and has, since then, played a very important role in medical imaging. There are different techniques available to simulate a process like OCT. It is possible to use Monte Carlo\cite{Kirillin2010,Periyasamy2016}, computational electrodynamics or ray tracing techniques\cite{Troiani2016}. Monte Carlo techniques rely on random sampling and can, in principle, be used to solve any problems having a probabilistic interpretation. Computational electrodynamics includes all the techniques that model, through approximations of Maxwell equations, the interaction of electromagnetic waves with physical objects and the environment. They work best when the wavelength of the electromagnetic wave considered is comparable with the smallest detail of the studied object. Examples of computational electrodynamics techniques are the method of moments, the boundary element method and the finite element method and FDTD method \cite{Hubing2008}. Ray tracing is a method that works best when the wavelength of the considered radiation is much smaller than the smallest detail of the studied object. Rays are advanced by a set distance and for each step the different properties of the ray (direction, intensity, wavelength, polarization) and its possible intersection with the studied objects are calculated.

\subsection{OCT for neural recording}
During the years, different techniques have been developed to record neural activity, varying in the level of invasiveness. Non invasive techniques -- e.g. EEG (Electro EncephaloGraphy, monitoring brain activity by recording electric signals from the surface of the scalp) or fMRI (functional Magnetic Resonance Imaging, measuring brain activity by detecting associated changes in blood flow) -- measure activity at a global level. On the other hand, invasive techniques -- e.g. penetrating microelectrodes -- are capable of a much higher spatial resolution. At this point in time, it is necessary to establish techniques that can give the best of the two worlds, providing both high resolution and low invasiveness. Optical recordings have been gaining momentum in the past 50 years and the advent of calcium sensitive dyes and genetically encoded indicators has revolutionised neural recordings. However both techniques exhibit either phototoxicity or very low brightness and this limits their usage in \textit{in vivo} experiments.

Research from the late '60s showed that change in scattered light from a nerve bundle during activity is of the order of ten parts per million and that the optical effect lasts for roughly a millisecond\cite{Cohen1968} and, in more recent years, scientists have been focussing on ways to obtain information on neural activity by observing changes in intrinsic optical properties of the neurons. As the use of OCT has already been proved useful for structural imaging of neural tissue\cite{Magnain2015,Boppart2003,Boppart1996}, we are working on using OCT to detect compound action potential in a peripheral nerve. The methods and simulation tool described herein has been developed to gain insights to this specific application.

\section{Method}

The FDTD method is one of the simplest of the computational electrodynamics techniques and it can solve a broad range of problems in a very accurate way. It was first developed by Yee \cite{Yee1966} in 1966, when the main applications of electromagnetic simulations were in the defence field, and from 1990 the interest in this technique has expanded to many different areas \cite{Taflove2005}.

FDTD is a grid-based differential numerical modelling method where both the spatial and temporal derivatives that appear in Maxwell's equations are discretised using a central-difference approximation. Central-difference approximations exist in several orders, this work uses a second order one:
\begin{equation}
\left.\frac{df(x)}{dx}\right|_{x=x_0}\approx \frac{f(x_0+\frac{h}{2})-f(x_0-\frac{h}{2})}{h}.
\label{eq:secondorder}
\end{equation}
It has to be noted that this approach provides an approximation of the value of the derivative of the function at $x_0$ but the function is sampled at the neighbouring points $x_0+h$ and $x_0-h$.

The algorithm used in this work is the Yee algorithm \cite{Yee1966}. Despite being the first version of FDTD, this algorithm is very robust and of easy implementation and it can be summarised as follows:
\begin{enumerate}
	\item Space and time are discretised -- for each point in space and time $(i,j,k,t)= (i\Delta x, j\Delta y, k\Delta z, q\Delta t)$ -- and the derivatives in Ampere's and Faraday's laws are replaced with finite differences.
	\item A new set of "updated equations" expressing the new fields in term of the past ones is obtained solving the difference equations.
	\item The magnetic field is evaluated.
	\item The electric field is evaluated.
	\item The previous three steps are repeated until the end of the simulation is reached.
\end{enumerate}
This algorithm can be applied to one, two and three-dimensional problems. The simulations presented in this paper are bi-dimensional as for both the samples considered the region of interest lays in the cross section.
\begin{figure}
	\centering
	\includegraphics[width=0.5\textwidth]{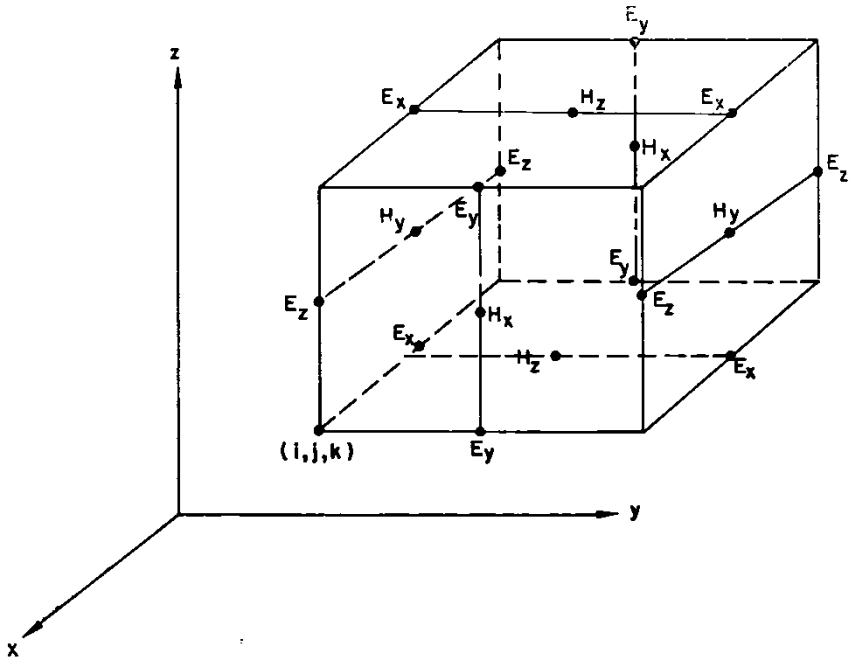}
	\caption{\textsc{Yee lattice.} The components of the electric field are in the middle of the edges and the components of the magnetic fields are in the centre of the faces \cite{Yee1966}.}
\end{figure}

\begin{figure}
	\centering
    \hfill
	\subfloat[\label{fig:spectrum500}]{\includegraphics[width=0.45\textwidth]{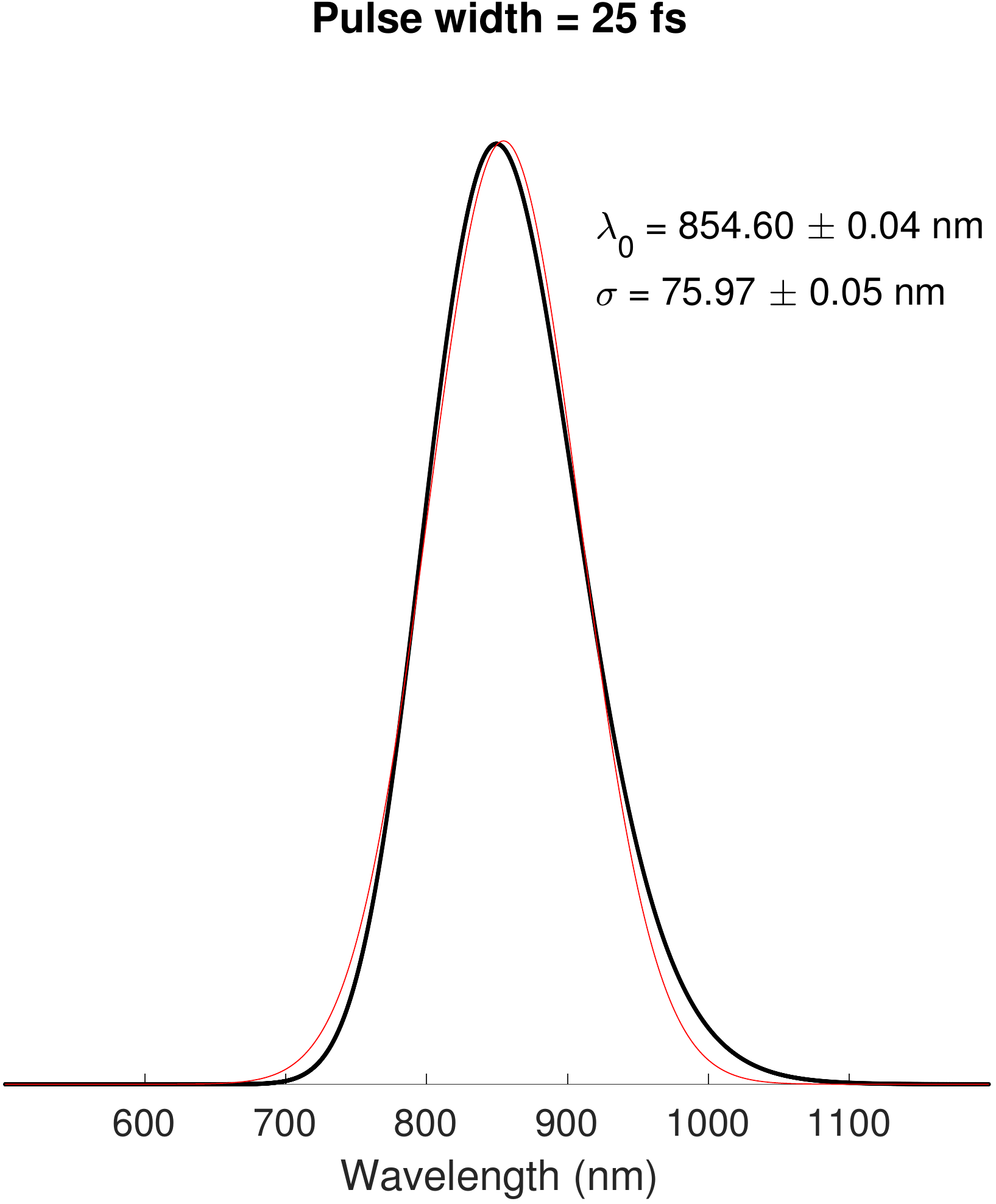}}\hfill
	\subfloat[\label{fig:spectrum3000}] {\includegraphics[width=0.45\textwidth]{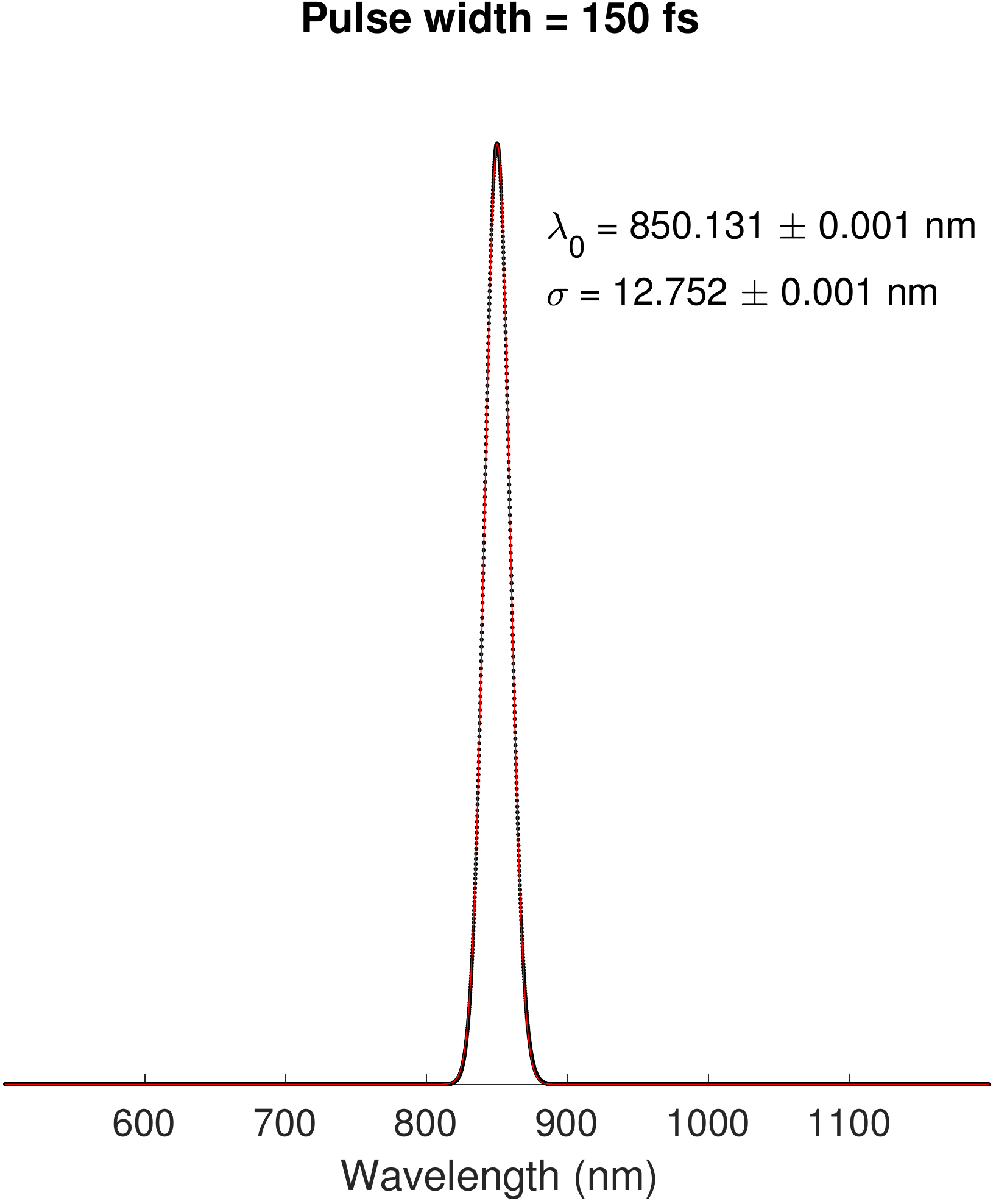}}\hfill
    \hfill
	\caption{Frequency spectrum for pulses of different lengths, fitted with a Gaussian curve. It is possible to notice that the broader spectrum is not a perfect Gaussian; this is due to the numerical dispersion being more pronounced in a source with a bigger frequency range.} \label{fig:spectra}
	
\end{figure}

One of the most interesting features of FDTD technique is that it allows to obtain results for a range of frequencies using a single simulation. It is thus possible to simulate the low coherence gate property of OCT by using, as the light source, a pulse which length in time is chosen to match the desired width of the frequency spectrum. \autoref{fig:spectra} shows the spectra obtained from pulses with a width of 25 fs and 150 fs and reported as a function of the wavelength.

For FDTD to be reliable and converge, the discretisation of the space needs to be at least an order of magnitude smaller than the wavelength. This means that at visible and near infra-red wavelengths, there is a limit on the dimension of the domain for the simulation to be carried out in a reasonable amount of time. Moreover, the temporal and spatial steps have to be in a relation such as the Courant number $S_c = \frac{c\Delta t}{\Delta x} \leq \frac{1}{\sqrt{D}}$, where c is the light speed in vacuum and D is the number of dimensions of the simulation. To this day there are a few FDTD models for OCT imaging, which -- while being optimised, parallelised and run on institutional clusters -- require more than a day to obtain the required scans \cite{Munro2015,Munro2016}. The aim of this model is to obtain the scans in a much shorter amount of time by simulating only the end of the two arms of the interferometer and computing the OCT signal in post processing.

\section{The model}
\label{sec:model}

Without loss of generality, the field is assumed to be polarised in the x-direction and presents a Gaussian shape. Assuming the light source is far from both ends of the interferometer, it is possible to represent the wave travelling towards both arms as following:
\begin{equation}
\overrightarrow{E} = E_0 \exp{\left(-\frac{(y-y_0)^2}{\sigma_y^2}\right)}\exp{\left(-\frac{(t-t_0)^2}{\sigma_t^2}\right)}\sin{\left(\frac{2\pi}{\lambda_0}\frac{c}{n}t\right)}\hat{x},
\label{eq:source}
\end{equation}
where $E_0$ is the amplitude of the incident field, $\lambda_0=850$ nm is the central wavelength and $n=1.3290$ is the refractive index of the water in which the sample is bathed at $850$ nm and $25^\circ$C\cite{Hale1973}.

The simulations reported in this paper have been run with both pulse lengths shown in \autoref{fig:spectra}. These two specific pulse lengths have been chosen because the one in Figure \ref{fig:spectrum500} represents a superluminescent diode (SLD) with a broadness of $\sigma \simeq 76 $ nm, which is quite common in the OCT community, while the one in Figure \ref{fig:spectrum3000} represents a diode with $\sigma \simeq 13$ nm, which corresponds to what can be typically found in low cost SLDs.

Two different samples have been studied using this computational model: a glass rod filled with sucrose solutions at different concentration and part of a myelinated nerve made of one single fascicles of axons (e.g. \textit{xenopus' laevis} sciatic nerve). In both cases the refractive indexes of the different layers of the simulation domain are considered to be constant over the range of frequencies considered, therefore the only dispersion in the simulations is the numerical one which is due to the discrete nature of the FDTD grid.

\subsection{Glass rod model}
\begin{figure}
\begin{center}
\includegraphics[width = 0.75\textwidth]{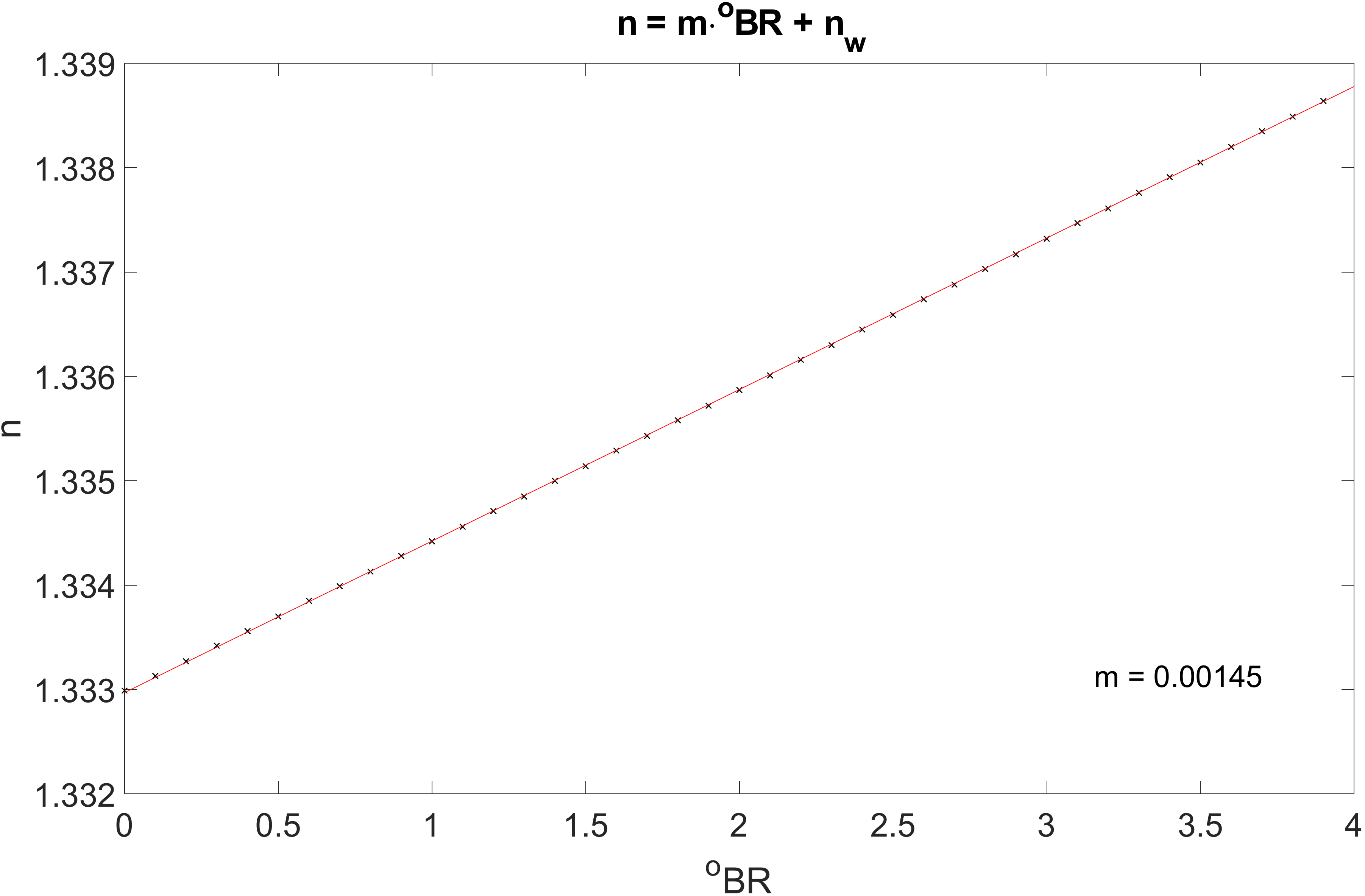}
\caption{Refractive index of a water-sucrose solution as a function of the sucrose concentration at 589.3 nm\cite{SugarBrix}.} \label{fig:brix}
\end{center}
\end{figure}

This model has been developed to obtain a result that could be easily verified experimentally. By changing the concentration of the water-sucrose solution inside the rods it is possible to obtain different values for its refractive index which can be computed using the BRIX scale\cite{SugarBrix} (one degree Brix is defined as 1 g of sucrose dissolved in 100 g of water). \autoref{fig:brix} shows that, at low concentrations, the refractive index of the solution is directly proportional to the concentration of sugar in the solution. The data used to obtain this value have been obtained for a wavelength of 589.3 nm and it has been assumed that to obtain the data for a different wavelength it is possible to translate rigidly the curve so that the intercept corresponds to the refractive index of water at the chosen wavelength. In the case of $\lambda = 850$ nm, this results in the curve becoming $n = m\cdot ^oBR + n_{w@850nm}$, where $m = 0.00145$ and $ n_{w@850nm} = 1.3290$.

\subsection{Nerve model}
\begin{figure}
\rotatebox{90}{
\begin{minipage}[c][0.99\textwidth][c]{0.99\textheight}
\hspace*{\fill}\includegraphics[width = 0.4\textheight]{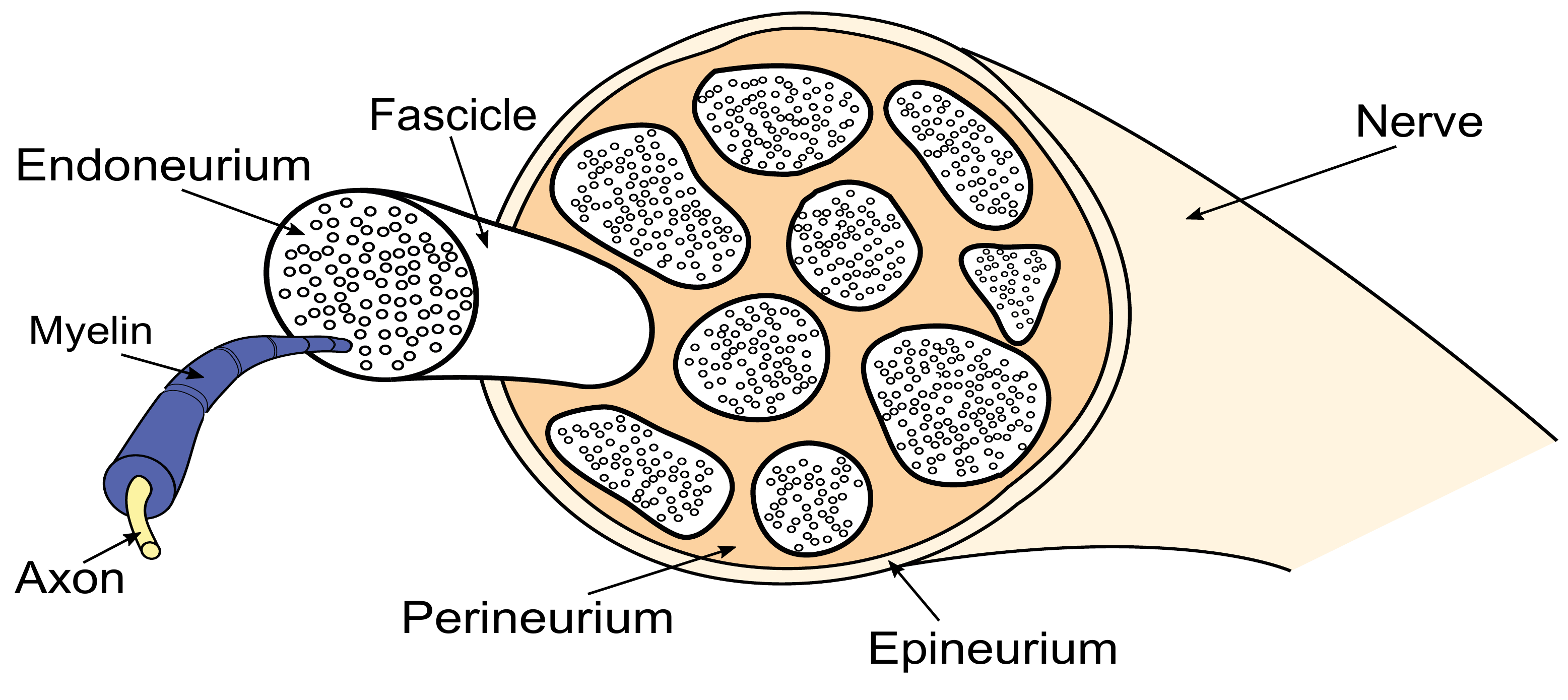}\hfill
\includegraphics[width = 0.2\textheight, angle = 90]{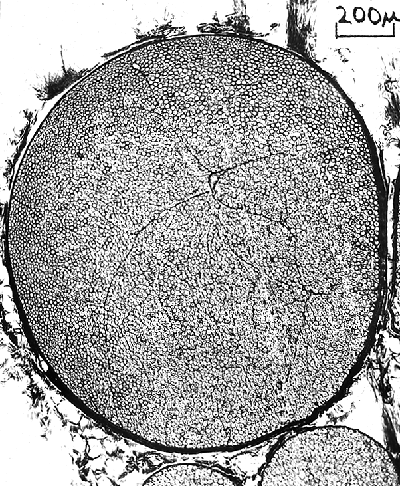}\hspace*{\fill}\\
\vspace*{\fill}
\hspace*{\fill}\includegraphics[width = 0.98\textheight]{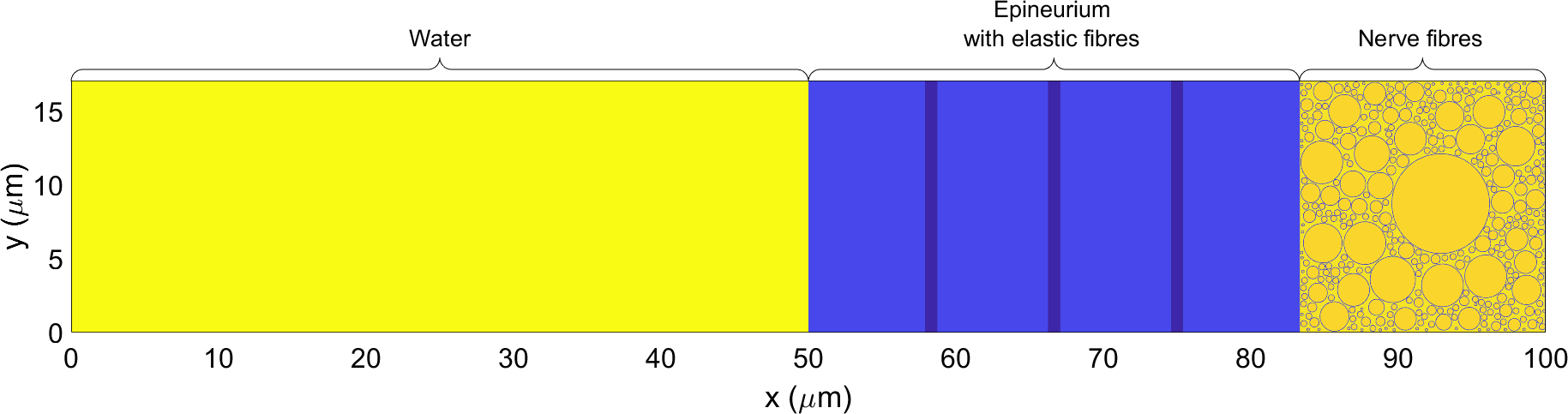}\hspace*{\fill}\\
\vspace*{\fill}
\hspace*{\fill}\begin{tabular}{c|c|c|c|c|c|c}
Water & Epineurium & Elastic fibres & Axons (inactive) & Axons (active) & Myelin sheath & Endoneurium\\ \hline
1.3290 & 1.4 & 1.41 & 1.338 & $1.338\cdot\left(1-10^{-6}\right)$& 1.4& 1.335\\
\end{tabular}\hspace*{\fill}\\
\caption{Top left: schematic structure of a nerve. Top right: \textit{Xenopus laevis'} sciatic nerve cross section\cite{McGill}. Bottom: simulation domain for the nerve model and the refractive index values for its different parts.\label{fig:nerve}}
\end{minipage}
}\end{figure}

The top left panel of \autoref{fig:nerve} shows the structure of a nerve and its different layers, while the top right panel shows a cross section of the nerve considered for this study, the \textit{Xenopus laevis}' sciatic nerve\cite{McGill}. This specific nerve is widely used in neuroscientific applications due to being made of one single fascicles, property that allows for a clearer signal in both the electrical and, considering the lack of multiple interfaces between different fascicles, the optical measurements.

The bottom panel shows the simulation domain considered for the nerve model. It is a domain of $100 \times 17 \mu m^2$ of which the first half is made of water and the second of the nerve and its fibres. In the \textit{Xenopus laevis'} sciatic nerve the outermost layer is a combination of perineurium and loose epineurium which are both dense connective tissue containing collagen fibrils, elastic fibres, small blood vessels and a variable amount of fat. This is modelled as a region (n = 1.4) that contains three smaller regions (n = 1.41) representing elastic fibres. Enclosed by the connective tissue layers are the axon fibres (n = 1.338 and n = $1.338\cdot(1-10^{-6})$ for the inactive and active fibres respectively) which are surrounded by myelin sheath (n = 1.4) and endoneurium tissue (n = 1.335).
Since the time an electromagnetic wave needs to travel a few centimetres in air and water is on the nanosecond scale, it is possible to simulate the scattering change undergone by neurons by running different simulations changing discretely the value of the refractive index of the sample.

\subsection{Post-processing}
\begin{figure}
\begin{center}
\includegraphics[width = 0.32\textwidth]{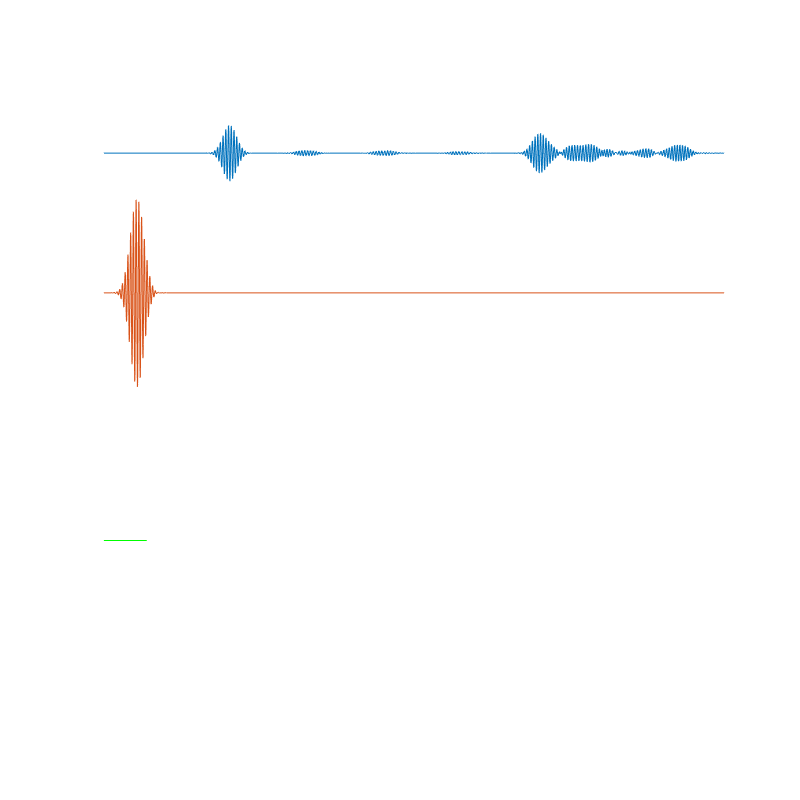}\includegraphics[width = 0.32\textwidth]{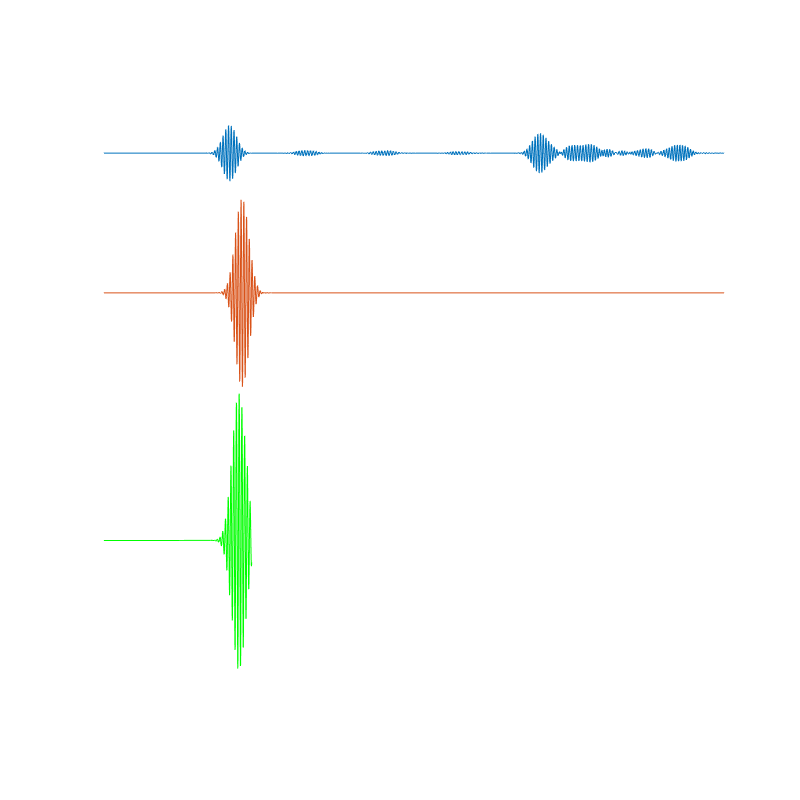}\includegraphics[width = 0.32\textwidth]{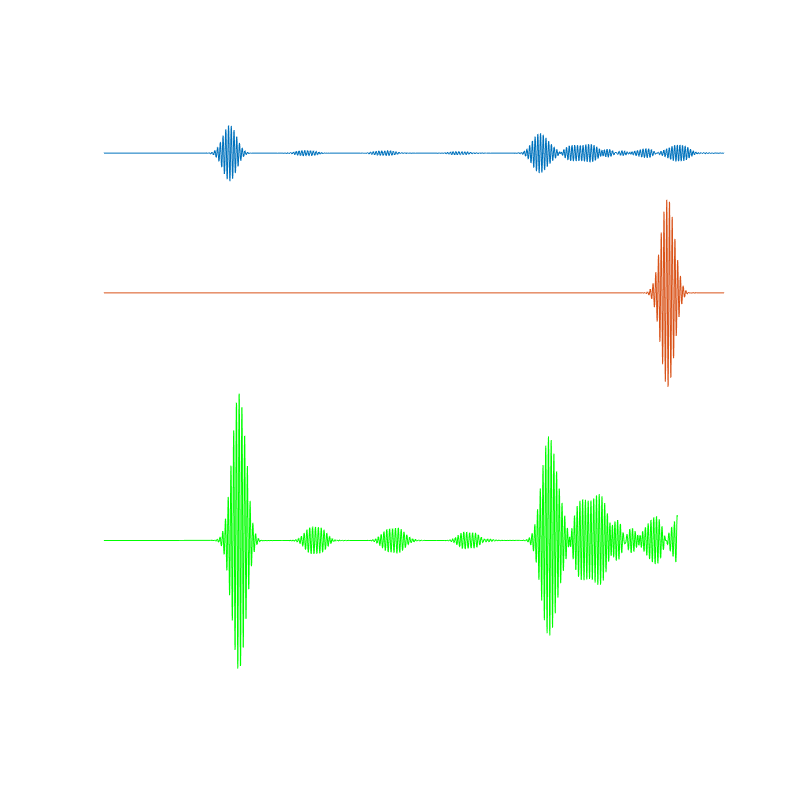}
\caption{Different stages of the post processing. In blue is the sample signal, in red the reference signal and in green the sum of the two.\label{fig:postprocessing}}
\end{center}
\end{figure}

As mentioned earlier, we have decided to remove from the simulation domain the whole splitter/combiner part and to obtain the interference between the reference and sample arms using an external software (Matlab R2017a). The FDTD simulations output the amplitude of the electric field in a preselected point every ten time steps. For each of the saved steps the signals obtained from the nerve and mirror simulations are summed and squared to obtain the intensity interference pattern (\autoref{fig:postprocessing} shows different stages of the process). The OCT signal is then computed as the envelope of the interference pattern. As this is a time domain simulation, each step in time corresponds to a spatial step and therefore the signal as a function of time can be converted in a signal as a function of tissue depth.

\section{Results}
\autoref{tab:parameters} shows the parameters used in the simulations reported in this paper.
\begin{table}
\centering
\resizebox{\textwidth}{!}{
\begin{tabular}{c|c|c|c}
Spatial step & Temporal step & Wavelength & Grid dimensions\\
\hline
$\Delta x = \Delta y = 8.5nm$ & $\Delta t = S_c \cdot \frac{\Delta x}{c} = 2\times 10^{-17}s$ & $\lambda_0 = 850 nm = 100\Delta x$ & $100\times 17 \mu m^2$
\end{tabular}}
\caption{Parameters used for the simulations.}\label{tab:parameters}
\end{table}

\begin{figure}
	\centering
	\includegraphics[width=0.8\textwidth]{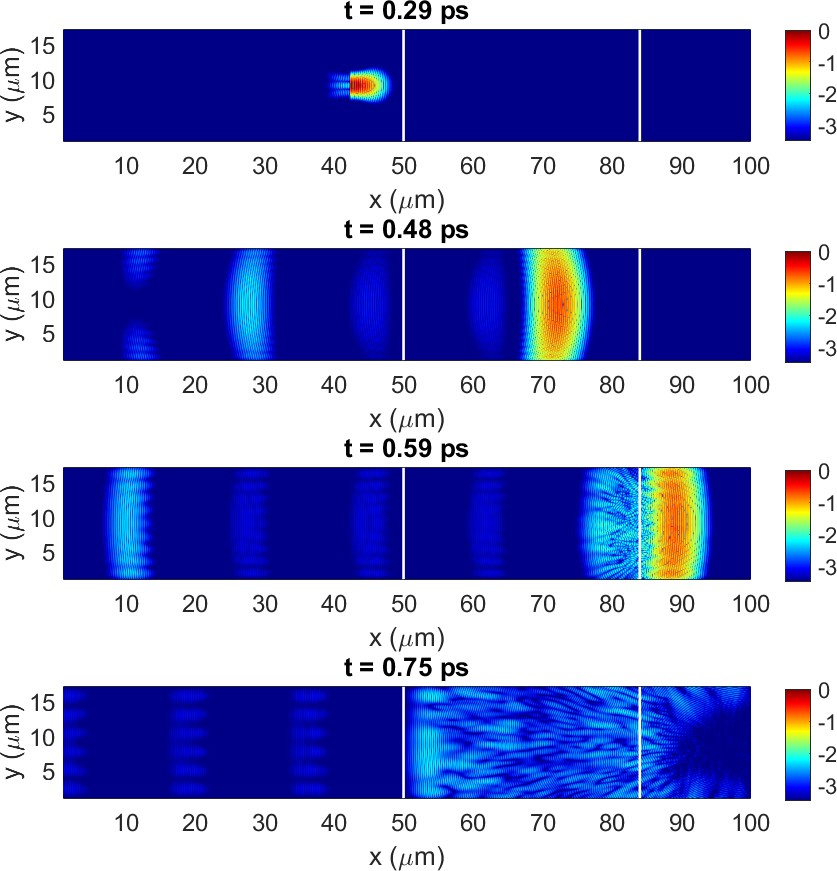}
    \caption{Colour map of the electric field amplitude at different points in time in the simulation. The first white vertical line represents the interface water-epineurium and the second the interface epineurium-nerve fibres. A video showing this can be seen in \url{https://vimeo.com/241209424}.}\label{fig:snapshots}
\end{figure}

\begin{figure}
	\centering
	\includegraphics[width=0.9\textwidth]{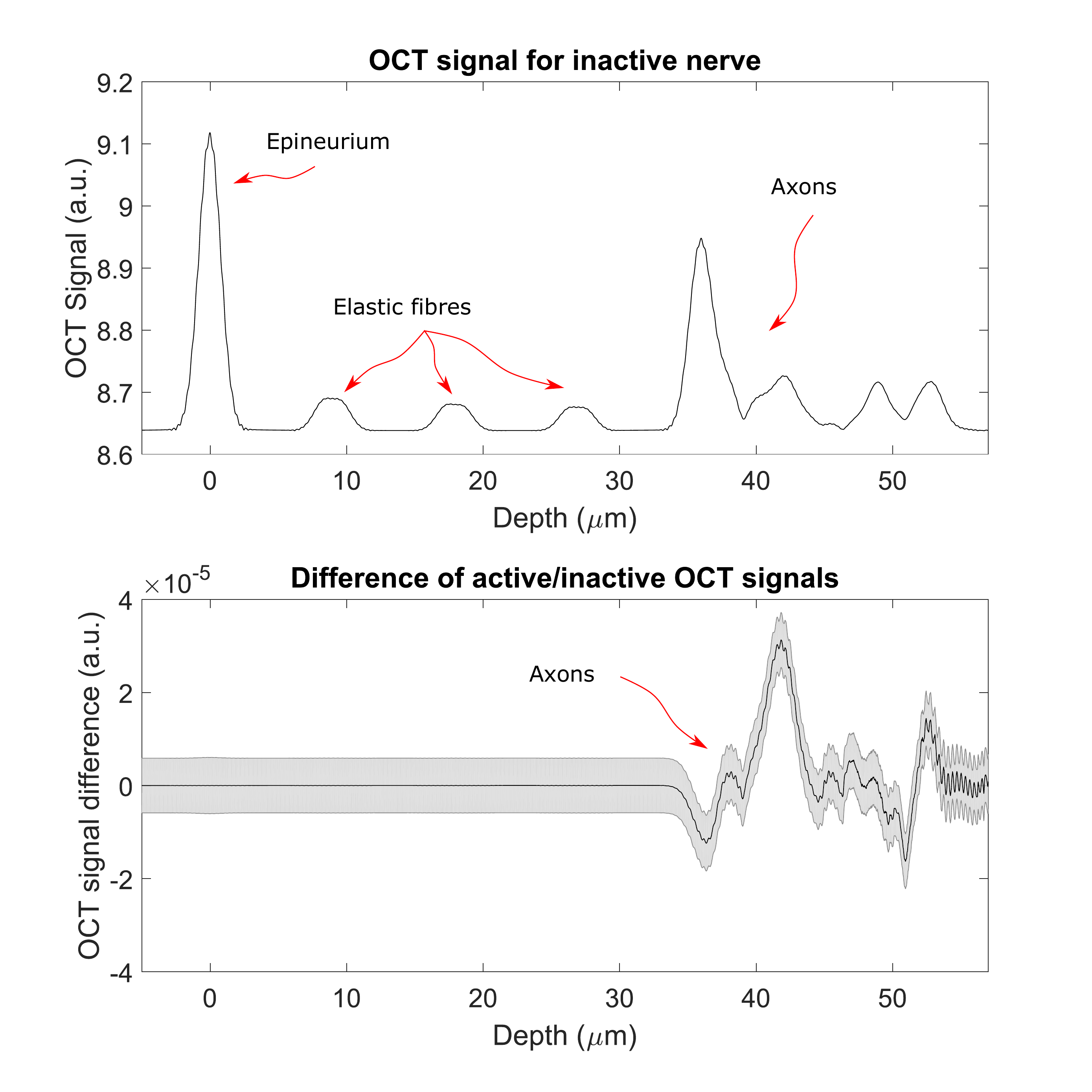}
    \caption{OCT signal for source with a $\sigma_\lambda \sim 13$ nm. The zero has been placed at the interface between water and epineurium and the grey shadowed area represents the error.}\label{fig:resultshighres}
\end{figure}

\begin{figure}
	\centering
	\includegraphics[width=0.75\textwidth]{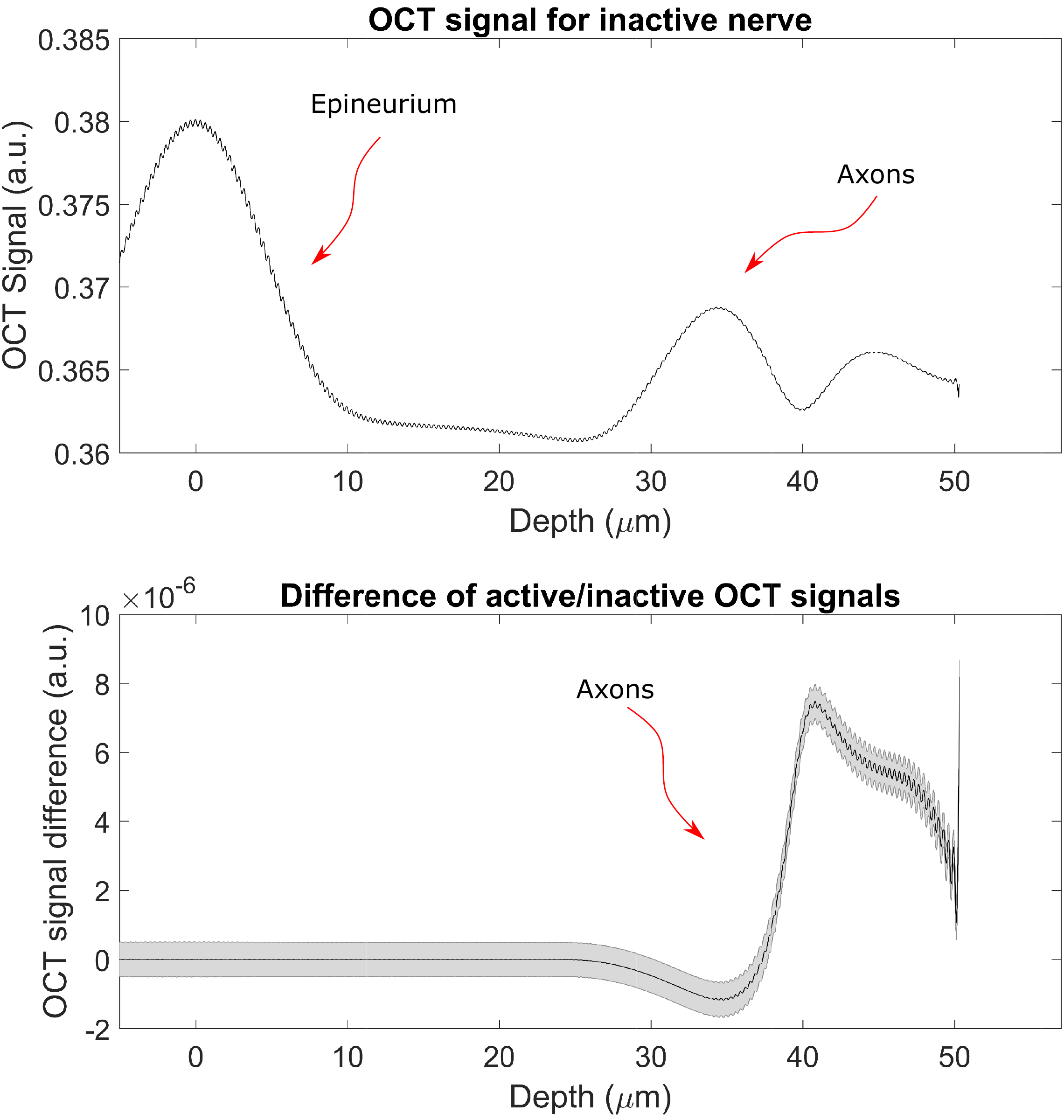}
    \caption{OCT signal for source with a $\sigma_\lambda \sim 76$ nm. The zero has been placed at the interface between water and epineurium and the grey shadowed area represents the error.}\label{fig:resultslowres}
\end{figure}

The simulations have been run with the two pulse lengths shown in \autoref{fig:spectra} for both models. After being post-processed, the signal assumes its final form where -- depending on the resolution -- it is possible to distinguish the different interfaces. The difference in resolution is visible by comparing \autoref{fig:resultshighres} and \autoref{fig:resultslowres}, obtained using the shorter and longer pulse respectively. \autoref{fig:snapshots} shows snapshots of the electric field amplitude taken at different times during the simulation.


\begin{figure}
	\centering
	\includegraphics[width=0.75\textwidth]{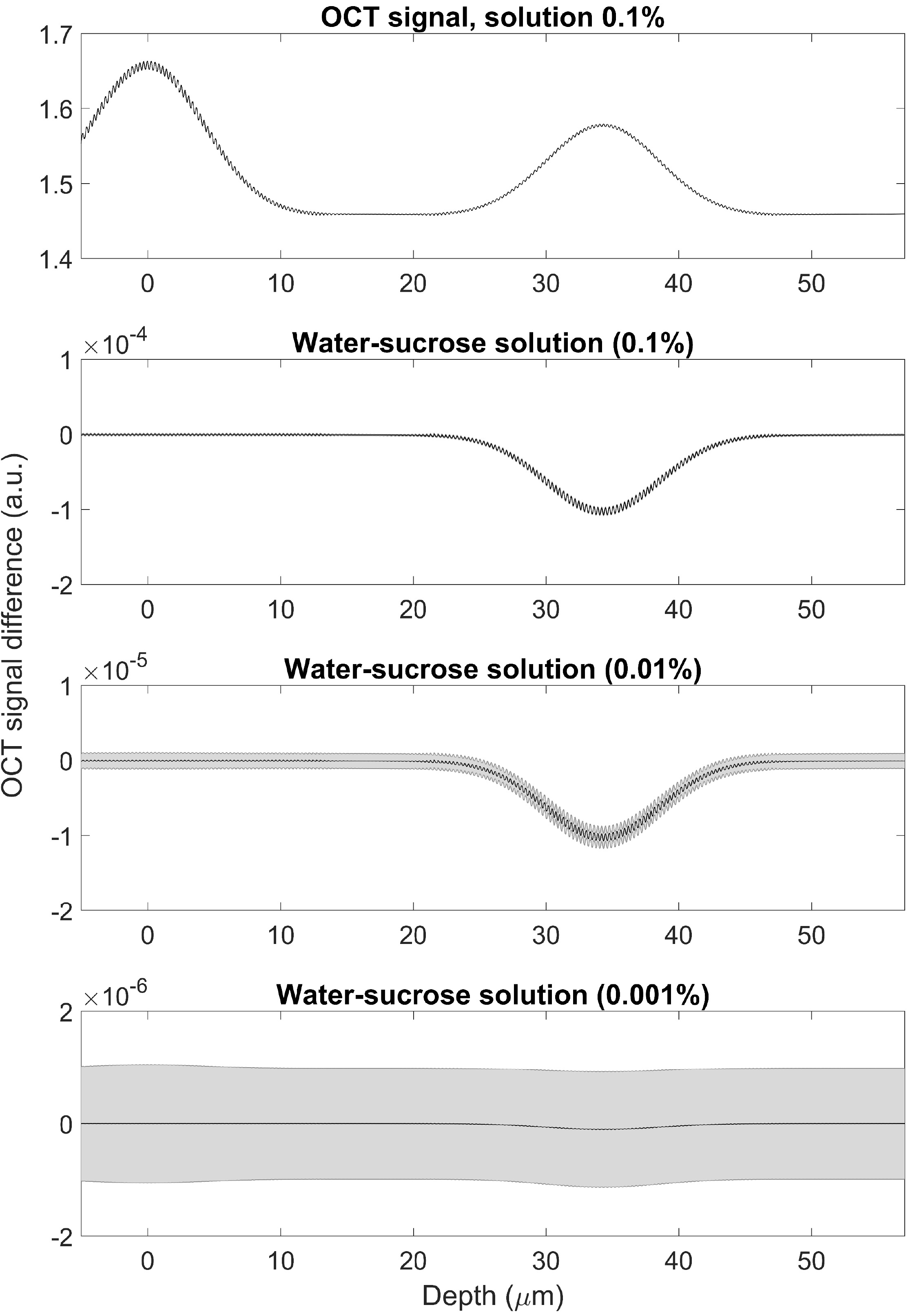}
    \caption{OCT signal for different concentrations of water-sucrose solutions and $\sigma_\lambda \sim 76$ nm. Top panel: signal obtained for the glass rods model with a water-sucrose solution at 0.1\%. Other panels: difference between the signal obtained from different concentrations (0.1\%, 0.01\% and 0.001\% respectively) and pure water. The grey shadowed area represents the error.}\label{fig:sugar}
\end{figure}

The top panel of \autoref{fig:sugar} shows the OCT signal obtained for the glass rod model with a water-sucrose solution at 0.1\%, the first peak being the result of the interface between water and glass and the second one between glass and solution. The other panels in the figure show the difference between the signal obtained from different concentrations (0.1\%, 0.01\% and 0.001\% respectively) and pure water.
\section{Discussion and conclusions}
The simulation of the studied region of the nerve has given the expected results: in \autoref{fig:resultshighres}, top panel, it is possible to see how the simulated OCT signal allows to distinguish between all the different regions of the nerve: first the interface between the Ringer solution layer and the nerve surface, the three elastic fibres inside the epineurium and the beginning of the nerve fibres region. In this last region the fibres cannot be detected at a single level because of resolution limitations (both in the axial and transverse directions). In the bottom panel the signal obtained from the simulation representing the active nerve has been subtracted to the one obtained from the simulation representing inactive nerve. It is clearly visible that the signal from the part of the nerves which refractive index is common to both simulations cancels perfectly, leaving only a signal in the axon part. The same thing can be noticed in the bottom panel of \autoref{fig:resultslowres}, but in this case the resolution of the scan is lower.

\begin{figure}
	\centering
	\includegraphics[width=0.75\textwidth]{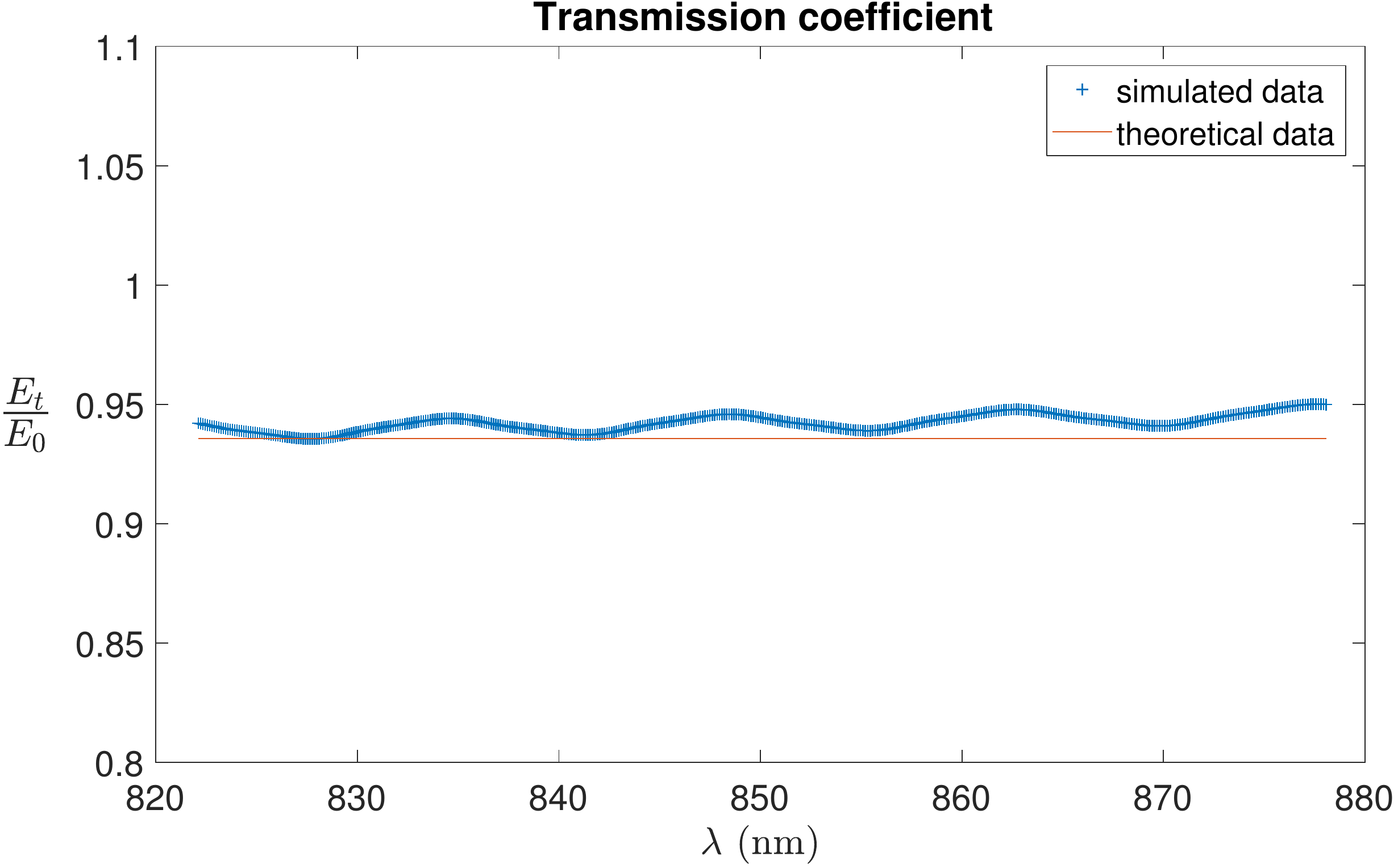}
    \caption{Transmission coefficient calculated theoretically and from the simulations.}\label{fig:coefficient}
\end{figure}

A water-sucrose solution at the 0.01\% appears to be a good approximation of the active nerve and an experiment has been planned to test the results obtained by the model before going into \textit{ex-vivo} studies of the nerve. \autoref{fig:coefficient} shows the transmission coefficient obtained for the interface water-glass in the simulations. The theoretical result, in absence of dispersion, is constant over the range of wavelengths considered; the simulated result is not perfectly compatible with the theoretical one and this is due to the unavoidable numerical dispersion that is intrinsic in the discreteness of the FDTD. There are various techniques that can be implemented to reduce the error introduced by the numerical dispersion, such as using a finer grid and implementing a fourth order approximation for Maxwell's equations.

The data obtained from the simulations, after post processing, provide an A-Line OCT scan. To obtain a B-scan it would be possible to create multiple simulation domains corresponding to contiguous part of the tissue considered and run a simulation for each of them. While it is true that splitting the simulation domain -- instead of having a domain that comprises the whole sample -- would result in an underestimation of the noise coming from scattering inside the bigger domain, it is also true that for biological tissues at near infra-red wavelengths light experiences primarily forward scattering\cite{Wilson1990}. Since multiple simulations can be run at the same time, this would allow for a remarkable reduction in computational time even for 3D images.

\section*{Funding}
This work has been supported by the UK Engineering and Physical Sciences Research Council (EPSRC).

\section*{Acknowledgements}
Portions of this work were presented at the European Conference of Biomedical Optics (ECBO) in June 2017, paper number 104160A. The FDTD code that served as a starting point for this work has been provided by John B. Schneider\cite{Schneider2010}. The code used to obtain the results showed in this paper can be found on \href{https://github.com/FTroiani/2D-FDTD-OCT}{GitHub/FTroiani}.
\end{document}